%% file: main.tex
\documentclass[10pt,letter,prl,twocolumn,hypertext]{revtex4-1}
\usepackage{ifthen}
\newboolean{uprightparticles}
\setboolean{uprightparticles}{false} 

\usepackage{amssymb}
\usepackage{amsfonts}
\usepackage{upgreek}
\usepackage{xspace}
\usepackage{color}
\usepackage{amsmath}
\usepackage{graphicx}
\usepackage{bm}
\usepackage{dcolumn}
\usepackage{hyperref}
\usepackage{bigstrut}

\newboolean{articletitles}
\setboolean{articletitles}{true} 

\newboolean{pdflatex}
\setboolean{pdflatex}{true} 

\input{lhcb-symbols-def}

\usepackage{mciteplus}
\begin{document}

\title{First evidence of direct {\boldmath $C\!P$} violation in charmless two-body decays of {\boldmath $B^0_s$} mesons}
\vspace*{1cm}
\author{\input{LHCb_authorlist}}
\collaboration{The LHCb collaboration\vspace{0.5cm}}
\begin{abstract}
Using a data sample corresponding to an integrated luminosity of 0.35 $\mathrm{fb}^{-1}$ collected by LHCb in 2011, we report the first evidence of $C\!P$ violation in the decays of $B^0_s$ mesons to $K^\pm \pi^\mp$ pairs, $A_{C\!P}(B^0_s \rightarrow K \pi)=0.27 \pm 0.08\,\mathrm{(stat)} \pm 0.02\,\mathrm{(syst)}$, with a significance of 3.3$\sigma$. Furthermore, we report the most precise measurement of $C\!P$ violation in the decays of $B^0$ mesons to $K^\pm \pi^\mp$ pairs, $A_{C\!P}(B^0 \rightarrow K\pi)=-0.088 \pm 0.011\,\mathrm{(stat)} \pm 0.008\,\mathrm{(syst)}$, with a significance exceeding 6$\sigma$.
\end{abstract}

\pacs{Valid PACS appear here}

\vspace*{-1cm}
\hspace{-9cm}
\mbox{\Large EUROPEAN ORGANIZATION FOR NUCLEAR RESEARCH (CERN)}

\vspace*{0.2cm}
\hspace*{-9cm}
\begin{tabular*}{16cm}{lc@{\extracolsep{\fill}}r}
\ifthenelse{\boolean{pdflatex}}
{\vspace*{-3.2cm}\mbox{\!\!\!\includegraphics[width=.14\textwidth]{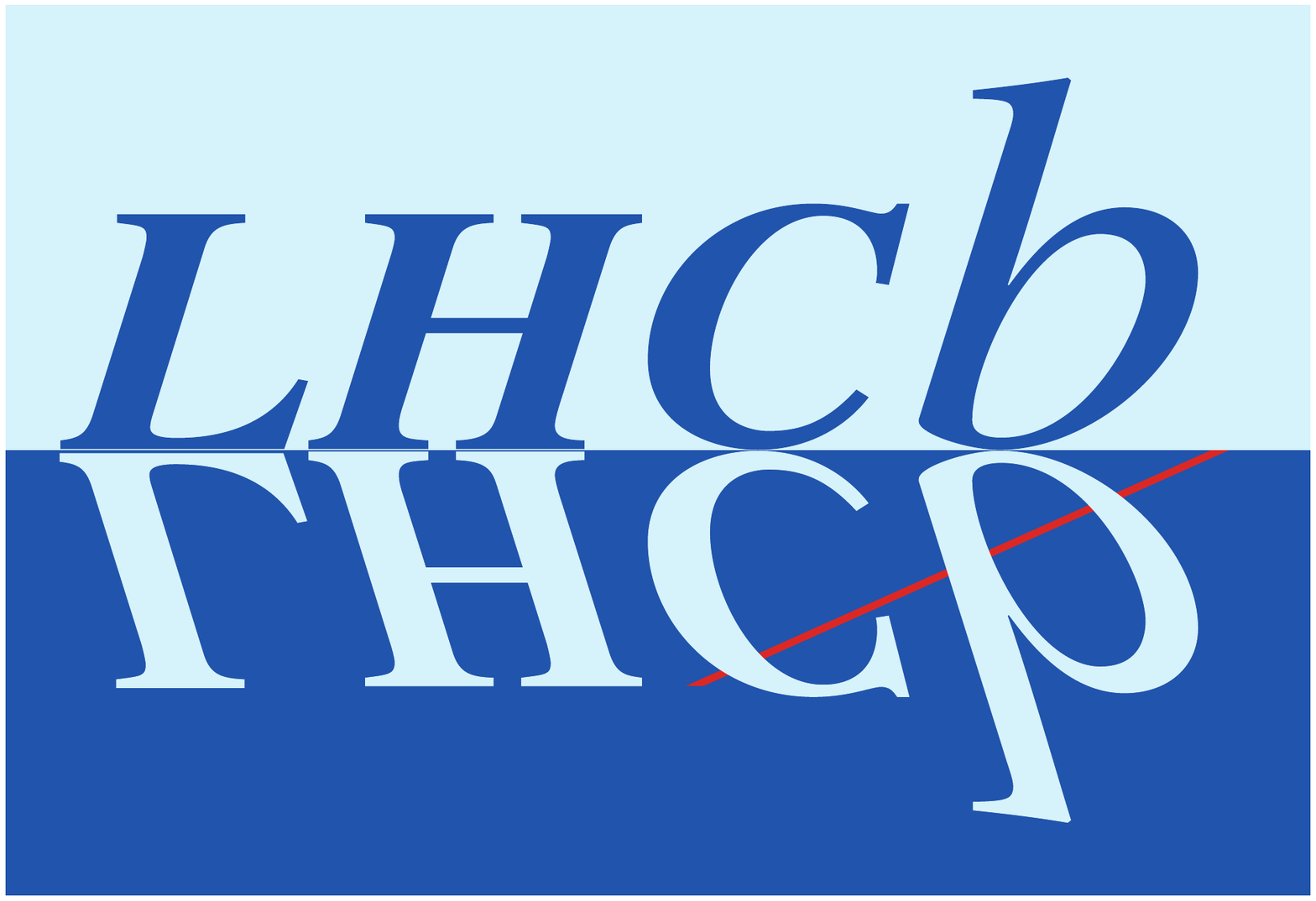}} & &}%
{\vspace*{-1.2cm}\mbox{\!\!\!\includegraphics[width=.12\textwidth]{lhcb-logo.eps}} & &}%
\\
 & & LHCb-PAPER-2011-029 \\
 & & CERN-PH-EP-2012-058 \\ 
 & & \today \\ 
\end{tabular*}
\vspace*{1cm}
\maketitle
\input{body}

\bibliographystyle{LHCb}
\bibliography{main}

\end{document}

%% file: lhcb-symbols-def.tex
\def\ux85 {UX85\xspace}

\ifthenelse{\boolean{uprightparticles}}%
{

 \def\PDelta      {\ensuremath{\Delta}\xspace}                 
 \def\PXi      {\ensuremath{\Xi}\xspace}                 
 \def\PLambda      {\ensuremath{\Lambda}\xspace}                 
 \def\PSigma      {\ensuremath{\Sigma}\xspace}                 
 \def\POmega      {\ensuremath{\Omega}\xspace}                 
 \def\PUpsilon      {\ensuremath{\Upsilon}\xspace}                 
 
 \def\PB      {\ensuremath{\mathrm{B}}\xspace}                 
                  
 \def\PD      {\ensuremath{\mathrm{D}}\xspace}

 \def\PK      {\ensuremath{\mathrm{K}}\xspace}

 \def\Pi      {\ensuremath{\mathrm{i}}\xspace}

}
{

 \mathchardef\PDelta="7101
 \mathchardef\PXi="7104
 \mathchardef\PLambda="7103
 \mathchardef\PSigma="7106
 \mathchardef\POmega="710A
 \mathchardef\PUpsilon="7107
                  
 \def\PB      {\ensuremath{B}\xspace}                 
                  
 \def\PD      {\ensuremath{D}\xspace}

 \def\PK      {\ensuremath{K}\xspace}

 \def\Pi      {\ensuremath{i}\xspace}

}

\def\kaon  {\ensuremath{\PK}\xspace}
  \def\Kbar  {\kern 0.2em\overline{\kern -0.2em \PK}{}\xspace}

\def\Kz    {\ensuremath{\kaon^0}\xspace}
\def\Kzb   {\ensuremath{\Kbar^0}\xspace}
\def\KzKzb {\ensuremath{\Kz \kern -0.16em \Kzb}\xspace}
\def\Kp    {\ensuremath{\kaon^+}\xspace}
\def\Km    {\ensuremath{\kaon^-}\xspace}

\def\KpKm  {\ensuremath{\Kp \kern -0.16em \Km}\xspace}

  \def\Dbar    {\kern 0.2em\overline{\kern -0.2em \PD}{}\xspace}
\def\D       {\ensuremath{\PD}\xspace}

\def\Dz      {\ensuremath{\D^0}\xspace}
\def\Dzb     {\ensuremath{\Dbar^0}\xspace}
\def\DzDzb   {\ensuremath{\Dz {\kern -0.16em \Dzb}}\xspace}
\def\Dp      {\ensuremath{\D^+}\xspace}
\def\Dm      {\ensuremath{\D^-}\xspace}

\def\DpDm    {\ensuremath{\Dp {\kern -0.16em \Dm}}\xspace}

  \def\Bbar    {\kern 0.18em\overline{\kern -0.18em \PB}{}\xspace}

  \def\Y#1S{\ensuremath{\PUpsilon{(#1S)}}\xspace}

\def\FL       {\ensuremath{F_{\mathrm{L}}}\xspace}
\def\AT#1     {\ensuremath{A_{\mathrm{T}}^{#1}}\xspace}           

\def\C#1      {\ensuremath{\mathcal{C}_{#1}}\xspace}                       
\def\Cp#1     {\ensuremath{\mathcal{C}_{#1}^{'}}\xspace}                    
\def\Ceff#1   {\ensuremath{\mathcal{C}_{#1}^{\mathrm{(eff)}}}\xspace}        
\def\Cpeff#1  {\ensuremath{\mathcal{C}_{#1}^{'\mathrm{(eff)}}}\xspace}       
\def\Ope#1    {\ensuremath{\mathcal{O}_{#1}}\xspace}                       
\def\Opep#1   {\ensuremath{\mathcal{O}_{#1}^{'}}\xspace}                    

\newcommand{\tev}{\ensuremath{\mathrm{\,Te\kern -0.1em V}}\xspace}
\newcommand{\gev}{\ensuremath{\mathrm{\,Ge\kern -0.1em V}}\xspace}
\newcommand{\mev}{\ensuremath{\mathrm{\,Me\kern -0.1em V}}\xspace}
\newcommand{\kev}{\ensuremath{\mathrm{\,ke\kern -0.1em V}}\xspace}
\newcommand{\ev}{\ensuremath{\mathrm{\,e\kern -0.1em V}}\xspace}
\newcommand{\gevc}{\ensuremath{{\mathrm{\,Ge\kern -0.1em V\!/}c}}\xspace}
\newcommand{\mevc}{\ensuremath{{\mathrm{\,Me\kern -0.1em V\!/}c}}\xspace}
\newcommand{\gevcc}{\ensuremath{{\mathrm{\,Ge\kern -0.1em V\!/}c^2}}\xspace}
\newcommand{\gevgevcccc}{\ensuremath{{\mathrm{\,Ge\kern -0.1em V^2\!/}c^4}}\xspace}
\newcommand{\mevcc}{\ensuremath{{\mathrm{\,Me\kern -0.1em V\!/}c^2}}\xspace}

\def\gsim{{~\raise.15em\hbox{$>$}\kern-.85em
          \lower.35em\hbox{$\sim$}~}\xspace}
\def\lsim{{~\raise.15em\hbox{$<$}\kern-.85em
          \lower.35em\hbox{$\sim$}~}\xspace}

\def\tell1  {TELL1\xspace}
\def\ukl1   {UKL1\xspace}



%% file: LHCb_authorlist.tex
\begin{center}
R.~Aaij$^{38}$, 
C.~Abellan~Beteta$^{33,n}$, 
B.~Adeva$^{34}$, 
M.~Adinolfi$^{43}$, 
C.~Adrover$^{6}$, 
A.~Affolder$^{49}$, 
Z.~Ajaltouni$^{5}$, 
J.~Albrecht$^{35}$, 
F.~Alessio$^{35}$, 
M.~Alexander$^{48}$
% ...

{\footnotesize \it
% ...
$ ^{5}$Clermont Universit\'{e}, Universit\'{e} Blaise Pascal, CNRS/IN2P3, LPC, Clermont-Ferrand, France\\
$ ^{6}$CPPM, Aix-Marseille Universit\'{e}, CNRS/IN2P3, Marseille, France\\
% ...
$ ^{33}$Universitat de Barcelona, Barcelona, Spain\\
$ ^{34}$Universidad de Santiago de Compostela, Santiago de Compostela, Spain\\
$ ^{35}$European Organization for Nuclear Research (CERN), Geneva, Switzerland\\
% ...
$ ^{38}$Nikhef National Institute for Subatomic Physics, Amsterdam, The Netherlands\\
% ...
$ ^{43}$H.H. Wills Physics Laboratory, University of Bristol, Bristol, United Kingdom\\
% ...
$ ^{48}$School of Physics and Astronomy, University of Glasgow, Glasgow, United Kingdom\\
$ ^{49}$Oliver Lodge Laboratory, University of Liverpool, Liverpool, United Kingdom\\
% ...
$ ^{n}$LIFAELS, La Salle, Universitat Ramon Llull, Barcelona, Spain\\
% ...
}
\end{center}

%% file: body.tex
The violation of $C\!P$ symmetry, \emph{i.e.}~the non-invariance of fundamental forces under the combined action of the charge conjugation ($C$) and parity ($P$) transformations, is well established in the $K^0$ and $B^0$ meson systems~\cite{Christenson:1964fg,Aubert:2001nu,Abe:2001xe,Nakamura:2010zzi}. Recent results from the LHCb collaboration have also provided evidence for $C\!P$ violation in the decays of $D^0$ mesons~\cite{Aaij:2011in}. Consequently, there now remains only one neutral heavy meson system, the $B^0_s$, where $C\!P$ violation has not yet been seen. All current experimental measurements of $C\!P$ violation in the quark flavor sector are well described by the Cabibbo-Kobayashi-Maskawa mechanism~\cite{Cabibbo:1963yz,Kobayashi:1973fv} which is embedded in the framework of the Standard Model (SM). However, it is believed that the size of $C\!P$ violation in the SM is not sufficient to account for the asymmetry between matter and antimatter in the Universe \cite{Hou:2008xd}, hence additional sources of $C\!P$ symmetry breaking are being searched for as manifestations of physics beyond the SM.

In this Letter we report measurements of direct $C\!P$ violating asymmetries in $B^0 \rightarrow K^+\pi^-$ and $B^0_s \rightarrow K^- \pi^+$ decays using data collected with the LHCb detector.
The inclusion of charge-conjugate modes is implied except in the asymmetry definitions.
$C\!P$ violation in charmless two-body $B$ decays could potentially reveal the presence of physics beyond the
SM~\cite{Fleischer:1999pa,Gronau:2000md,Lipkin:2005pb,Fleischer:2007hj,Fleischer:2010ib},
and has been extensively studied at the $B$ factories and at the Tevatron~\cite{Aubert:2008sb,Belle:2008zza,Aaltonen:2011qt}.
The direct $C\!P$ asymmetry in the $B^0_{(s)}$ decay rate to the final state $f_{(s)}$, with $f=K^+\pi^-$ and $f_s=K^-\pi^+$, is defined as
\begin{equation}
A_{C\!P} = \Phi \! \left[\Gamma \! \left(\overline{B}^0_{(s)} \rightarrow \bar{f}_{(s)} \right)\!\!,\,\Gamma \! \left(B^0_{(s)} \rightarrow f_{(s)}\right)\right]\!\!,\label{eq:acp}
\end{equation}
where $\Phi[X,\,Y] = (X-Y)/(X+Y)$ and $\bar{f}_{(s)}$ denotes the charge-conjugate of $f_{(s)}$.

LHCb is a forward spectrometer covering the pseudo-rapidity range $2<\eta<5$, designed to perform flavor physics measurements at the LHC.
A detailed description of the detector can be found in Ref.~\cite{Alves:2008zz}.
The analysis is based on $pp$ collision data collected in the first half of 2011 at a center-of-mass energy of $7\!$~\tev, corresponding to an integrated luminosity of $0.35~\mathrm{fb}^{-1}$.
The polarity of the LHCb magnetic field is reversed from time to time in order to partially cancel the effects of instrumental charge asymmetries,
and about $0.15~\mathrm{fb}^{-1}$ were acquired with one polarity and $0.20~\mathrm{fb}^{-1}$ with the opposite polarity.

The LHCb trigger system comprises a hardware trigger followed by a High Level Trigger (HLT) implemented in software. 
The hadronic hardware trigger selects high transverse energy clusters in the hadronic calorimeter.
A transverse energy threshold of 3.5~\gev has been adopted for the data set under study.
The HLT first selects events with at least one large transverse momentum track characterized by a large impact parameter, and then uses algorithms to reconstruct $D$ and $B$ meson decays.
Most of the events containing the decays under study have been acquired by means of a dedicated two-body HLT selection. To discriminate between signal and background events, this trigger selection imposes requirements on: the quality of the online-reconstructed tracks ($\chi^2$ per degree of freedom), their transverse momenta ($p_\mathrm{T}$) and their impact parameters ($d_\mathrm{IP}$, defined as the distance between the reconstructed trajectory of the track and the $pp$ collision vertex); the distance of closest approach of the decay products of the $B$ meson candidate ($d_\mathrm{CA}$), its transverse momentum ($p_\mathrm{T}^B$), its impact parameter ($d_\mathrm{IP}^B$) and the decay time in its rest frame ($t_{\pi\pi}$, calculated assuming the decay into $\pi^+\pi^-$). Only $B$ candidates within the $\pi\pi$ invariant mass range 4.7--5.9~\gevcc are accepted. The $\pi\pi$ mass hypothesis is conventionally chosen to select all charmless two-body $B$ decays using the same criteria.

Offline selection requirements are subsequently applied. Two sets of criteria have been optimized with the aim of minimizing the expected uncertainty either on $A_{C\!P}(B^0 \rightarrow K\pi)$ or on $A_{C\!P}(B^0_s \rightarrow K \pi)$. In addition to more selective requirements on the kinematic variables already used in the HLT, two further requirements on the larger of the transverse momenta and of the impact parameters of the daughter tracks are applied.
A summary of the two distinct sets of selection criteria is reported in Table~\ref{tab:selectioncuts}. In the case of $B^0_s \rightarrow K\pi$ decays a tighter selection is needed
because the probability for a $b$ quark to decay as $B^0_s \rightarrow K\pi $ is about 14 times smaller than that to decay as $B^0 \rightarrow K\pi$~\cite{Aaltonen:2008hg}, and consequently a stronger rejection of combinatorial background is required. 
The two samples passing the event selection are then subdivided into different final states
using the particle identification (PID) provided by the two ring-imaging Cherenkov (RICH) detectors.
Again two sets of PID selection criteria are applied: a loose set optimized for the measurement of $A_{C\!P}(B^0 \rightarrow K\pi)$ and a tight set for that of $A_{C\!P}(B^0_s \rightarrow K\pi)$.

\begin{table}
\caption{Summary of selection criteria adopted for the measurement of $A_{C\!P}(B^0 \rightarrow K\pi)$ and $A_{C\!P}(B^0_s \rightarrow K \pi)$.}
\begin{centering}
\begin{tabular}{lccc}
\hline
\hline  
Variable & $A_{C\!P}(B^0 \rightarrow K\pi)$ \bigstrut & $A_{C\!P}(B^0_s \rightarrow K \pi)$ \\
\hline
Track quality $\chi^2$/ndf & $<3$ & $<3$ \\
Track $p_\mathrm{T}\,[\textrm{Ge\kern -0.1em V\!/}c]$ & $>1.1$ & $>1.2$ \\
Track $d_\mathrm{IP}\,[\mathrm{mm}]$ & $>0.15$ & $>0.20$ \\
$\mathrm{max}(p_\mathrm{T}^{K},\, p_\mathrm{T}^{\pi})\,[\textrm{Ge\kern -0.1em V\!/}c]$ & $>2.8$ & $>3.0$ \\
$\mathrm{max}(d_\mathrm{IP}^{K},\,d_\mathrm{IP}^{\pi})\,[\mathrm{mm}]$ & $>0.3$ & $>0.4$ \\
$d_\mathrm{CA}$ $[\mathrm{mm}]$ & $<0.08$ & $<0.08$ \\
$p_\mathrm{T}^{B}\,[\textrm{Ge\kern -0.1em V\!/}c]$ & $>2.2$ & $>2.4$ \\
$d_\mathrm{IP}^{B}\,[\mathrm{mm}]$ & $<0.06$ & $<0.06$\\
$t_{\pi\pi}\,[\textrm{ps}]$ & $>0.9$ & $>1.5$ \\
\hline
\hline
\end{tabular}
\end{centering}
\label{tab:selectioncuts}
\vspace{-0.3cm}
\end{table}

\begin{figure*}[t]
\begin{center}
\begin{tabular}{cc}
\includegraphics[width=0.43\textwidth]{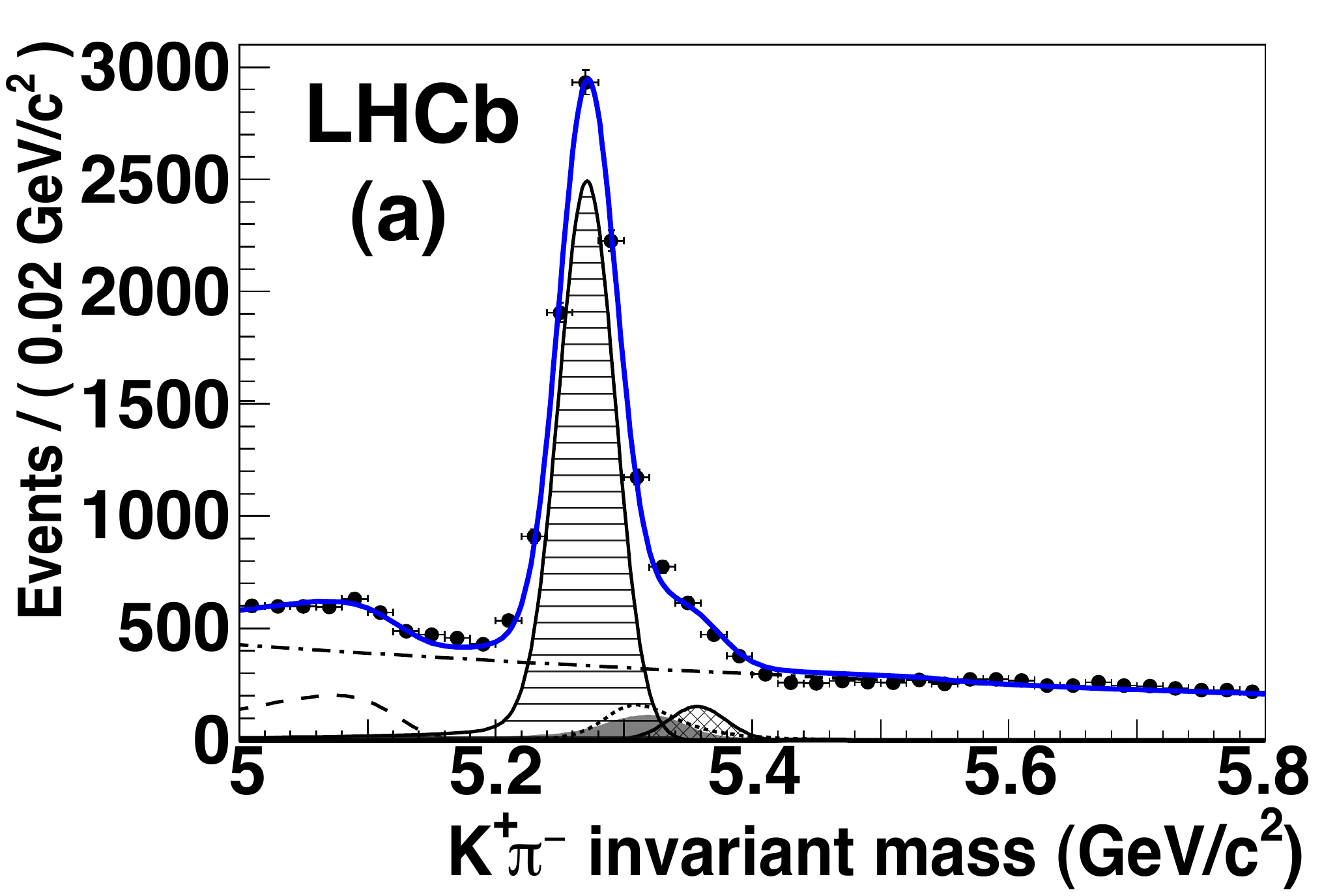} & \includegraphics[width=0.43\textwidth]{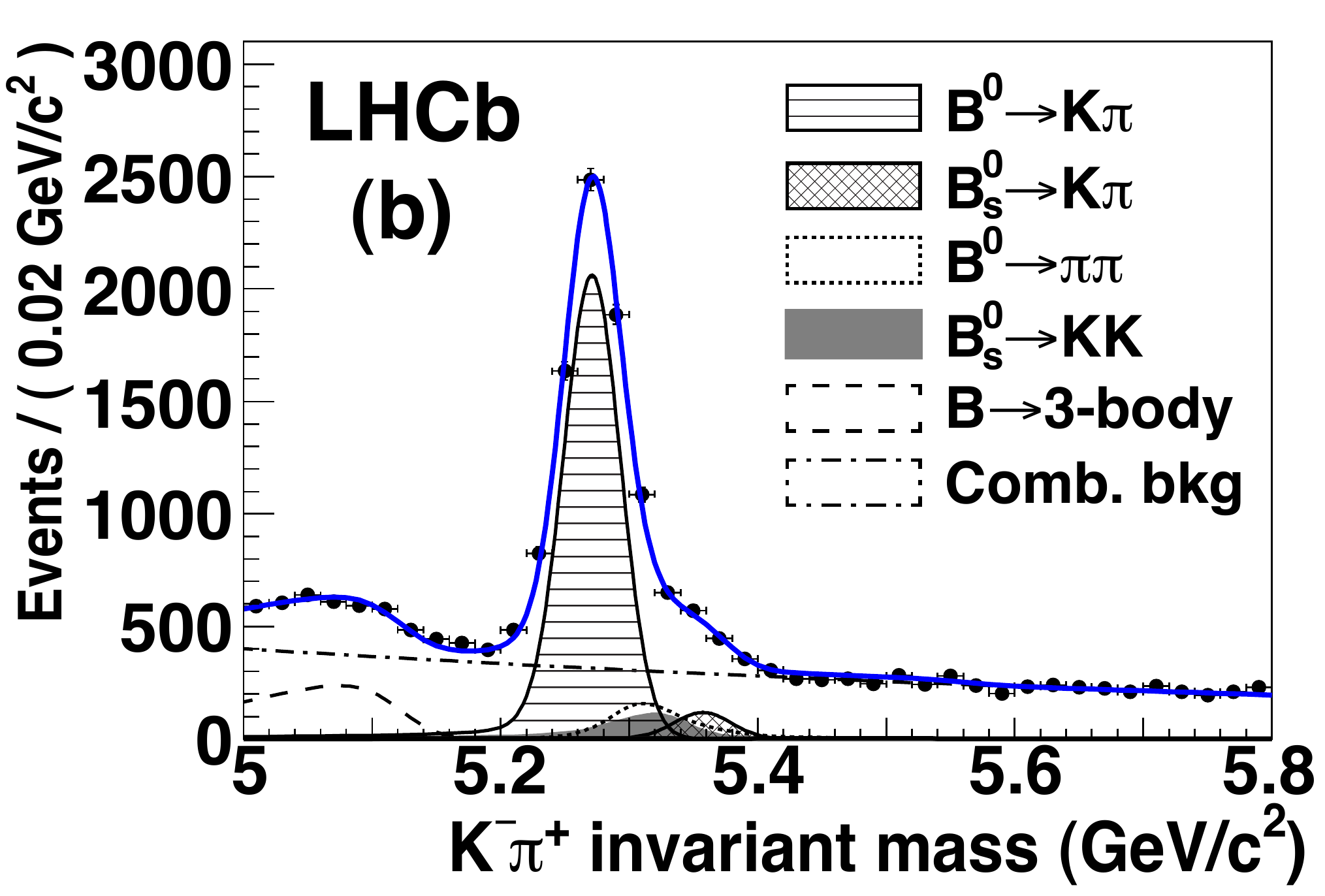}\\
\includegraphics[width=0.43\textwidth]{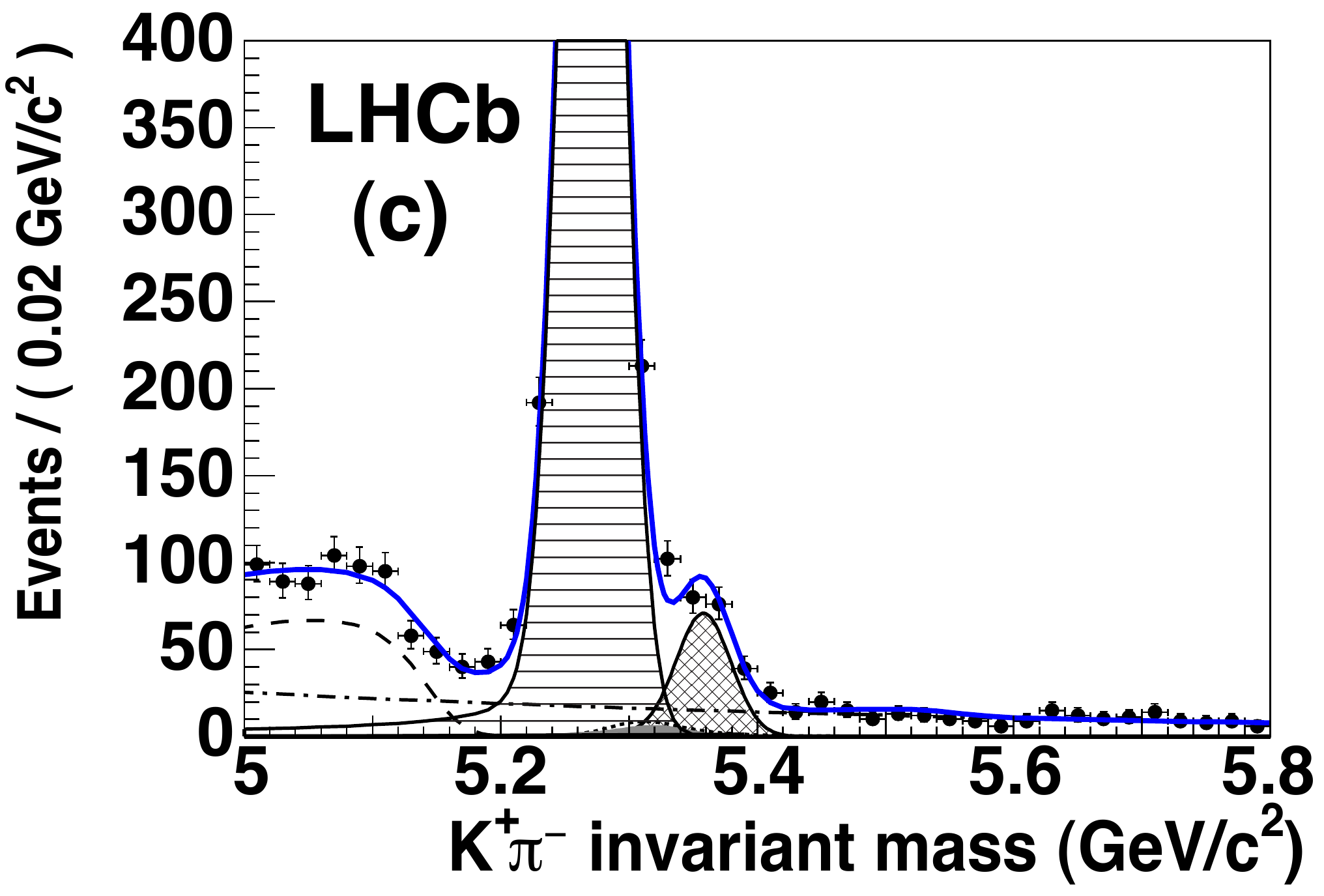} & \includegraphics[width=0.43\textwidth]{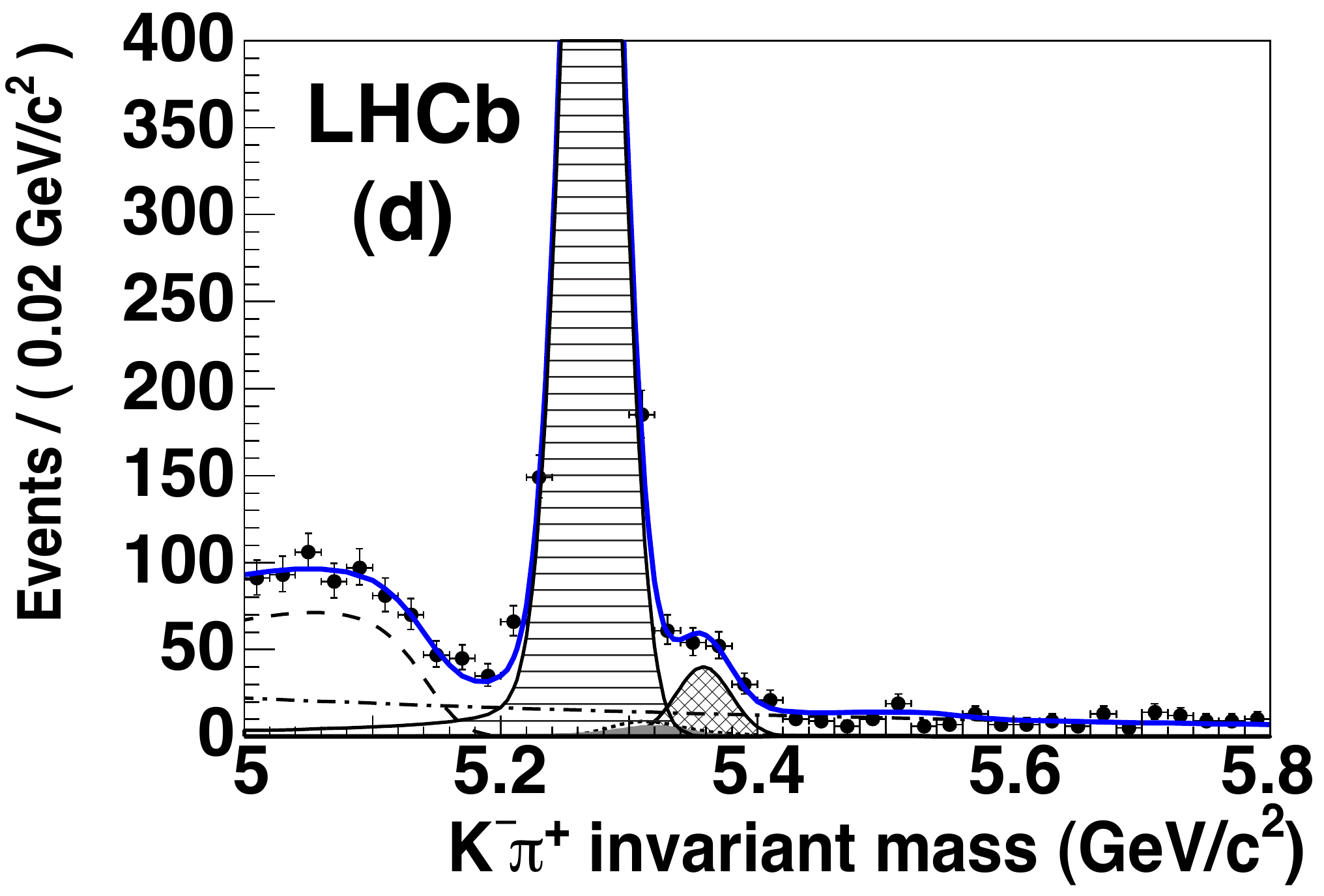}
\end{tabular}
\vspace{-0.5cm}
\end{center}
\caption{Invariant $K\pi$ mass spectra obtained using the event selection adopted for the best sensitivity on (a, b) $A_{C\!P}(B^0 \rightarrow K\pi)$ and (c, d) $A_{C\!P}(B^0_s \rightarrow K\pi)$. Plots (a) and (c) represent the $K^+\pi^-$ invariant mass whereas plots (b) and (d) represent the $K^-\pi^+$ invariant mass. The results of the unbinned maximum likelihood fits are overlaid. The main components contributing to the fit model are also shown.}
\label{fig:bd2kpifit_bd2kpi2}
\end{figure*}

To estimate the background from other two-body $B$ decays with a misidentified pion or kaon (cross-feed background), the relative efficiencies of the RICH PID selection criteria must be determined.
The high production
rate of charged $D^{*}$ mesons at the LHC and the kinematic characteristics of the
$D^{*+} \rightarrow D^0(K^-\pi^+) \pi^+$ decay chain
make such events an appropriate calibration sample for the PID of kaons and pions.
In addition, for calibrating the response of the RICH system for protons, a sample
of $\Lambda \rightarrow p \pi^-$ decays is used.
PID information is not used to select either sample, as the selection
of pure final states can be realized by means of kinematic criteria alone.
The production and decay kinematics of the $D^0 \rightarrow K^-\pi^+$ and $\Lambda \rightarrow p \pi^-$
channels differ from those of the $B$ decays under study. Since the RICH PID information is momentum dependent, the distributions obtained from calibration samples are reweighted
according to the momentum distributions of $B$ daughter tracks observed in data.

Unbinned maximum likelihood fits to the $K\pi$ mass spectra of the selected events are performed. 
The $B^0 \rightarrow K\pi$ and $B^0_s \rightarrow K\pi$ signal components are described by single Gaussian functions convolved with a function which describes the effect of final state radiation on the mass lineshape~\cite{Baracchini:2005wp}. The background due to partially reconstructed three-body $B$ decays is parameterized by means of an ARGUS function~\cite{Albrecht:1989ga} convolved with a Gaussian resolution function. The combinatorial background is modeled by an exponential and the shapes of the cross-feed backgrounds, mainly due to $B^0 \rightarrow \pi^+\pi^-$ and $B^0_s \rightarrow K^+K^-$ decays with one misidentified particle in the final state, are obtained from Monte Carlo simulations. The $B^0 \rightarrow \pi^+\pi^-$ and $B^0_s \rightarrow K^+K^-$ cross-feed background yields are determined from fits to the $\pi^+\pi^-$ and $K^+K^-$ mass spectra respectively, using events selected by the same offline selection as the signal and taking into account the appropriate PID efficiency factors. The $K^+\pi^-$ and $K^-\pi^+$ mass spectra for the events passing the two offline selections are shown in Fig.~\ref{fig:bd2kpifit_bd2kpi2}.

From the two mass fits we determine respectively the signal yields $N(B^0 \rightarrow K\pi)=13\hspace{0.5mm}250\pm 150$ and $N(B^0_s \rightarrow K\pi)=314 \pm 27$, as well as
the raw yield asymmetries $A_\mathrm{raw}(B^0 \rightarrow K\pi)=-0.095 \pm 0.011$ and $A_\mathrm{raw}(B^0_{s} \rightarrow K\pi)=0.28 \pm 0.08$, where the uncertainties are statistical only. In order to determine the $C\!P$ asymmetries from the observed raw asymmetries, effects induced by the detector acceptance and event reconstruction, as well as due to strong interactions of final state particles with
the detector material, need to be taken into account. Furthermore, the possible presence of a $B^0_{(s)}-\overline{B}^0_{(s)}$ production asymmetry must also be considered.
The $C\!P$ asymmetry is related to the raw asymmetry by $A_{C\!P}=A_\mathrm{raw} - A_\Delta$, where the correction $A_\Delta$ is defined as
\begin{equation}
A_\Delta (B^0_{(s)} \rightarrow K\pi) =\zeta_{d(s)}A_\mathrm{D}(K\pi)+\kappa_{d(s)} A_\mathrm{P}(B^0_{(s)}),
\end{equation}
where $\zeta_{d}=1$ and $\zeta_{s}=-1$, following the sign convention for $f$ and $f_s$ in Eq.~(\ref{eq:acp}).
The instrumental asymmetry $A_\mathrm{D}(K\pi)$ is given in terms of the detection efficiencies $\varepsilon_\mathrm{D}$ of the charge-conjugate final states by
$A_\mathrm{D}(K\pi) = \Phi[\varepsilon_\mathrm{D}(K^-\pi^+),\,\varepsilon_\mathrm{D}(K^+\pi^-)]$, and the production asymmetry $A_\mathrm{P}(B^0_{(s)})$ is defined in terms of the $\overline{B}^0_{(s)}$ and $B^0_{(s)}$ production rates, $R(\overline{B}^0_{(s)})$ and $R(B^0_{(s)})$, as $A_\mathrm{P}(B^0_{(s)}) = \Phi[R(\overline{B}^0_{(s)}),\,R(B^0_{(s)})]$. The factor $\kappa_{d(s)}$ takes into account dilution due to neutral $B^0_{(s)}$ meson mixing, and is defined as
\begin{equation}
\kappa_{d(s)} \!=\! \frac{\int_0^\infty \! e^{-\Gamma_{d(s)} t} \! \cos \!\left( \Delta m_{d(s)} t \right )\! \varepsilon(B^0_{(s)} \! \rightarrow \! K\pi;\,t)\mathrm{d}t}{\int_0^\infty \! e^{-\Gamma_{d(s)} t} \! \cosh \!\left( \frac{\Delta \Gamma_{d(s)}}{2} t \right )\! \varepsilon(B^0_{(s)} \! \rightarrow \! K\pi;\,t)\mathrm{d}t},
\end{equation}
where $\varepsilon(B^0 \rightarrow K\pi;\,t)$ and $\varepsilon(B^0_s \rightarrow K\pi;\,t)$ are the acceptances as functions of the decay time for the two reconstructed decays.
To calculate $\kappa_d$ and $\kappa_s$ we assume that $\Delta\Gamma_d=0$ and we use the world averages for $\Gamma_d$, $\Delta m_d$, $\Gamma_s$, $\Delta m_s$ and $\Delta\Gamma_s$~\cite{Nakamura:2010zzi}. The shapes of the acceptance functions are parameterized using signal decay time distributions extracted from data. We obtain $\kappa_d = 0.303 \pm 0.005$ and $\kappa_s = -0.033 \pm 0.003$, where the uncertainties are statistical only. In contrast to $\kappa_d$, the factor $\kappa_s$ is small, owing to the large $B^0_s$ oscillation frequency, thus leading to a negligible impact of a possible production asymmetry of $B^0_s$ mesons on the corresponding $C\!P$ asymmetry measurement.

The instrumental charge asymmetry $A_\mathrm{D}(K\pi)$ can be expressed in terms of two distinct contributions $A_\mathrm{D}(K\pi)=A_\mathrm{I}(K\pi)+\alpha(K\pi) A_\mathrm{R}(K\pi)$, where $A_\mathrm{I}(K\pi)$ is an asymmetry due to the different strong interaction cross-sections with the detector material of $K^+\pi^-$ and $K^-\pi^+$ final state particles, and $A_\mathrm{R}(K\pi)$ arises from the possible presence of a reconstruction or detection asymmetry. The quantity $A_\mathrm{I}(K\pi)$ does not change its value by reversing the magnetic field, as the difference in the interaction lengths seen by the positive and negative
particles for opposite polarities is small. By contrast, $A_\mathrm{R}(K\pi)$ changes its sign
when the magnetic field polarity is reversed.
The factor $\alpha(K\pi)$ accounts for different signal yields in the data sets with opposite polarities, due to the different values of the corresponding integrated luminosities and to changing trigger conditions in the course of the run. It is estimated by using the yields of the largest decay mode, \emph{i.e.} $B^0 \rightarrow K\pi$, determined from the mass fits applied to the two data sets separately. We obtain $\alpha(K\pi) = \Phi[ N ^{\rm up} (B^0\rightarrow K\pi),\,N ^{\rm down} (B^0\rightarrow K\pi) ] = -0.202 \pm 0.011$, where ``up'' and ``down'' denote the direction of the main component of the dipole field.

\begin{figure}[t]
\begin{center}
\begin{tabular}{cc}
\includegraphics[width=0.23\textwidth]{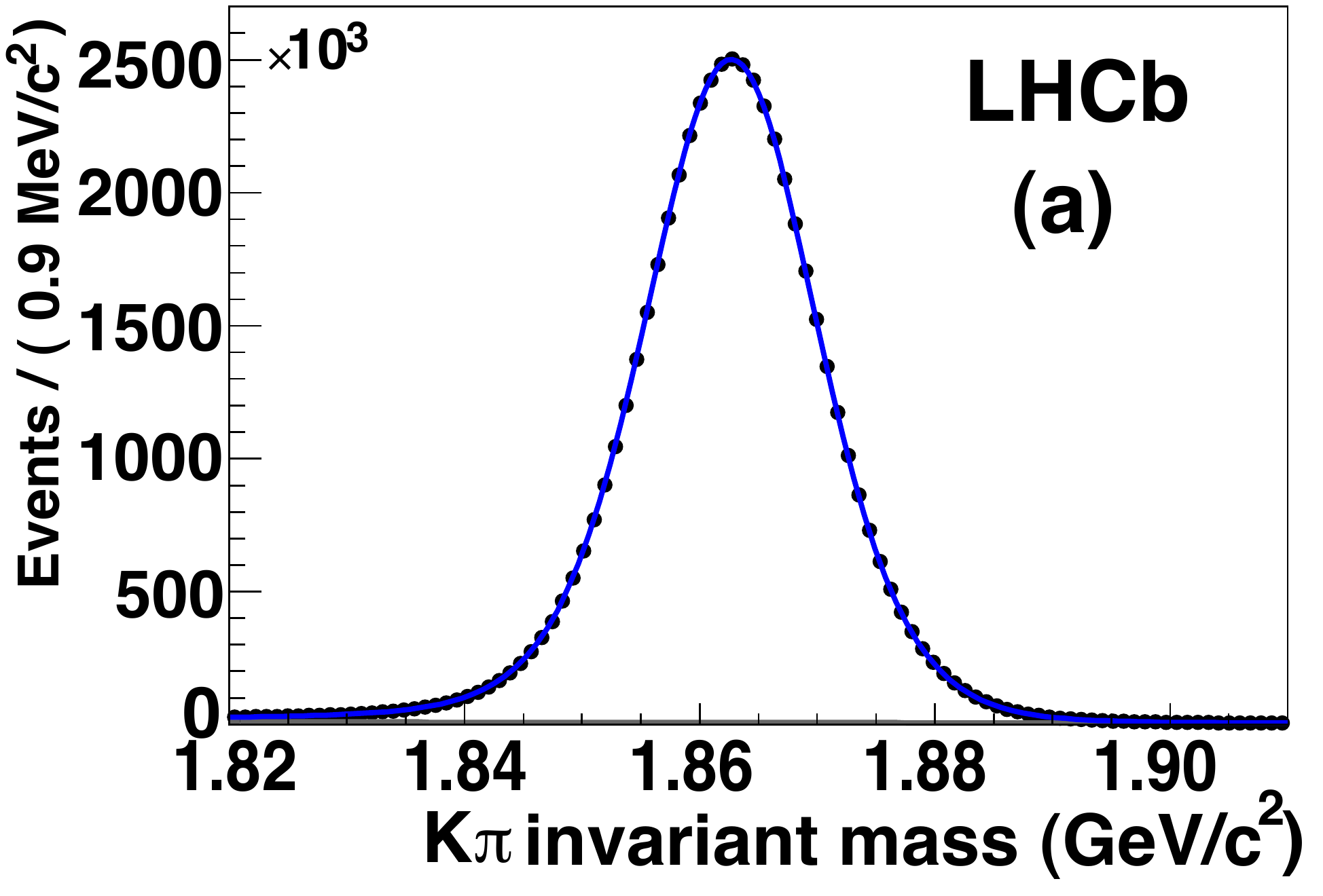}&
\includegraphics[width=0.23\textwidth]{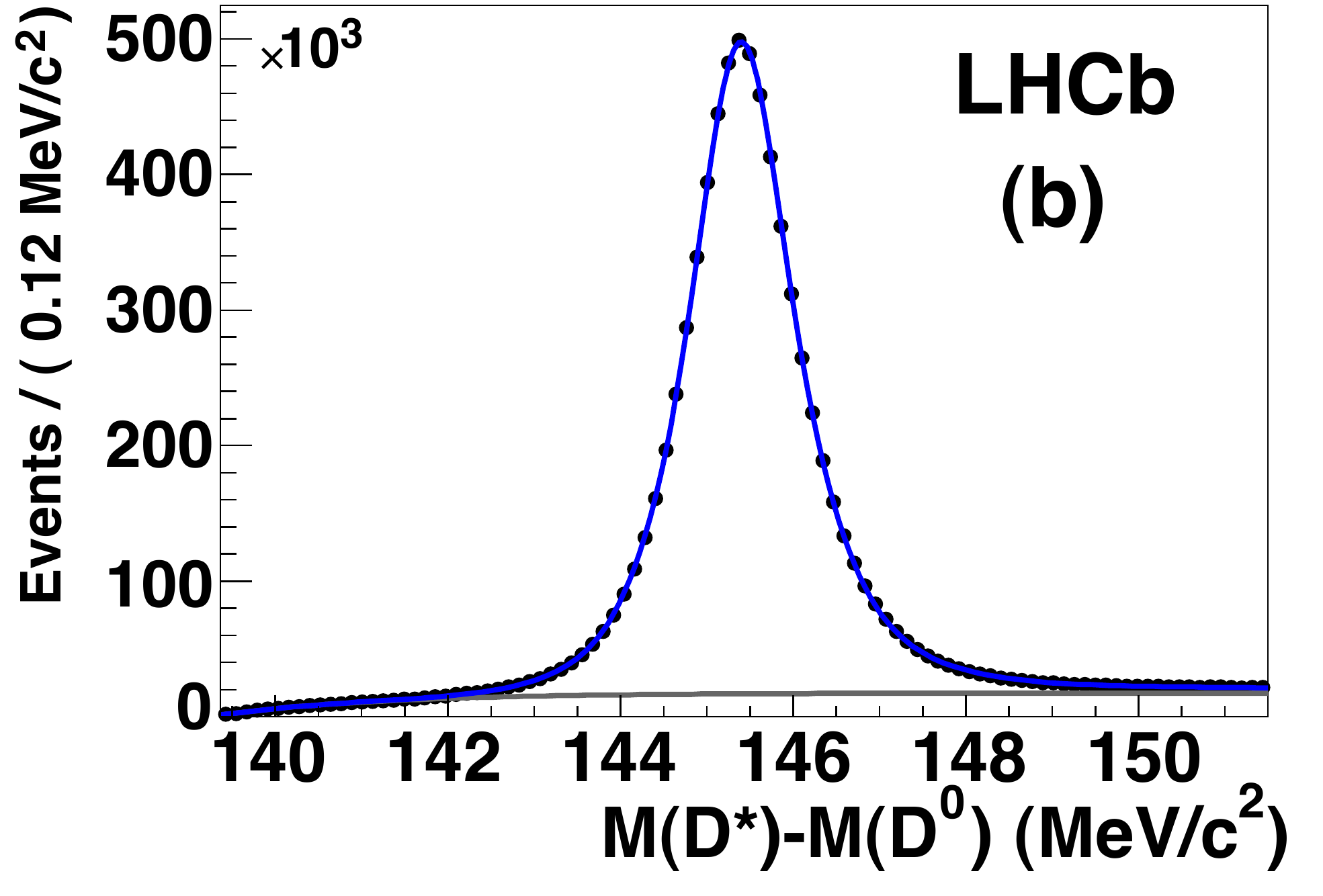}\\
\includegraphics[width=0.23\textwidth]{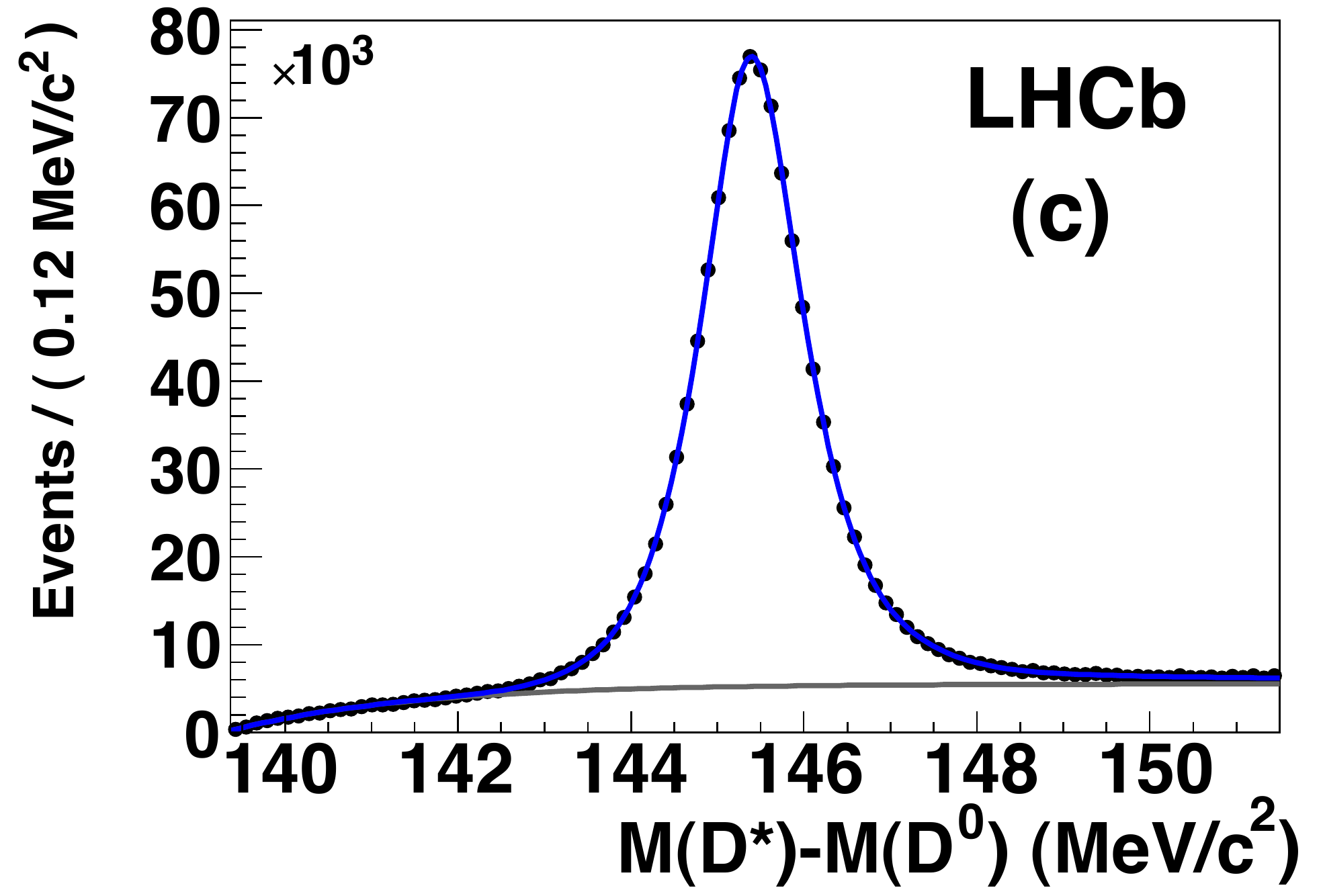}&
\includegraphics[width=0.23\textwidth]{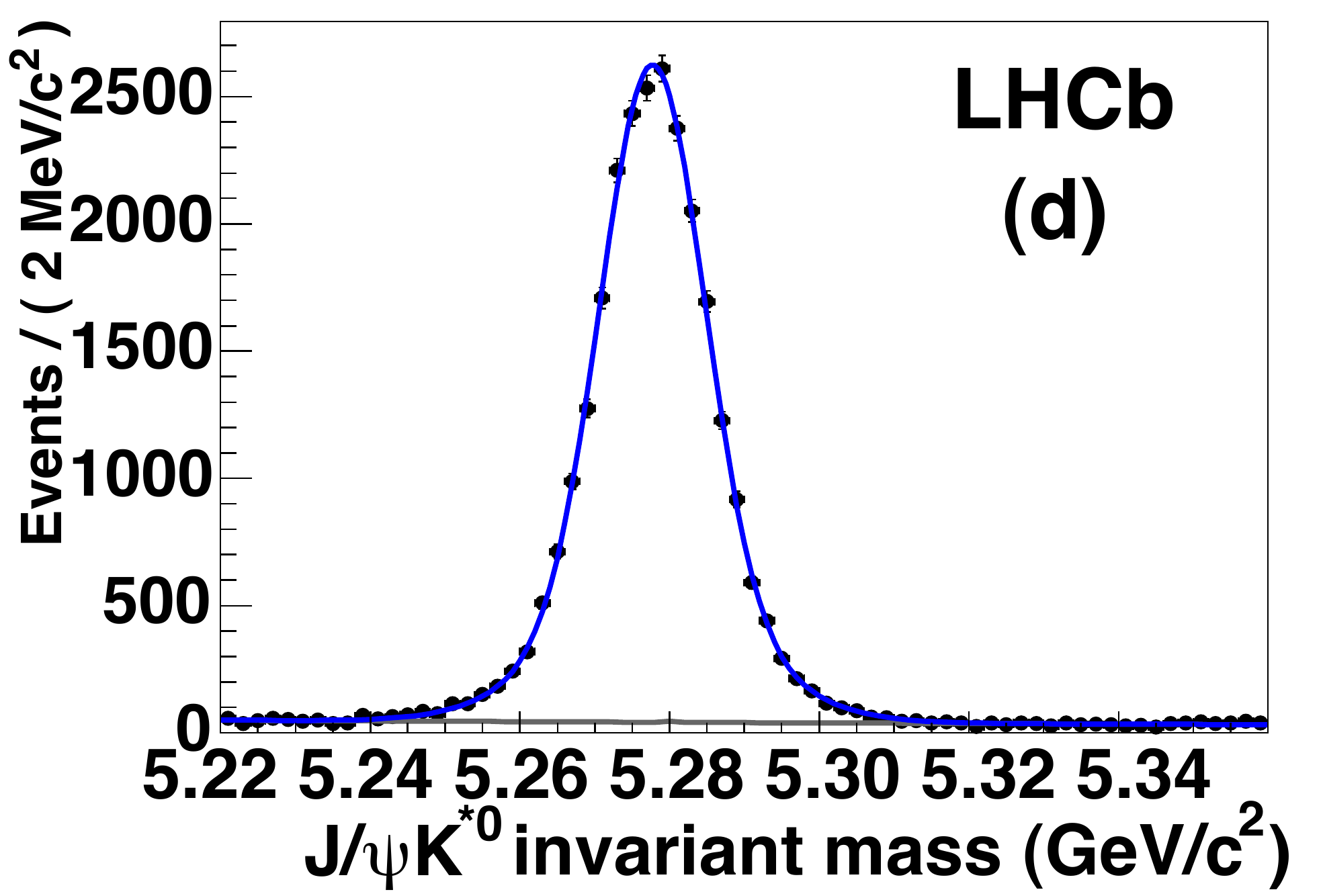} 
\end{tabular}
\vspace{-0.5cm}
\end{center}
\caption{Distributions of the invariant mass or invariant mass difference of (a) $D^{0}\rightarrow K^{-}\pi^{+}$, (b) $D^{*+}\rightarrow D^{0}(K^{-}\pi^{+})\pi^{+}$, (c) $D^{*+}\rightarrow D^{0}(K^{-}K^{+})\pi^{+}$ and (d) $B^{0}\rightarrow J/\psi(\mu^+\mu^-) K^{*0}(K^+\pi^-)$. The results of the maximum likelihood fits are overlaid.}
\label{fig:controlsamples}
\vspace{-0.3cm}
\end{figure}

The instrumental asymmetries for the final state $K\pi$ are measured from data using
large samples of tagged $D^{*+} \rightarrow D^0(K^-\pi^+)\pi^+$ and $D^{*+} \rightarrow D^0(K^-K^+)\pi^+$ decays, and untagged $D^0 \rightarrow K^-\pi^+$ decays.
The combination of the integrated raw asymmetries of
all these decay modes is necessary to disentangle the various contributions to the raw asymmetries of each mode, notably including the $K\pi$ instrumental asymmetry as well as that of the pion from the $D^{*+}$ decay, and the production asymmetries of the $D^{*+}$ and $D^0$ mesons.
In order to determine the raw asymmetry of the $D^{0}\rightarrow K\pi$ decay, a maximum likelihood fit to the
$K^{-}\pi^{+}$ and $K^{+}\pi^{-}$ mass spectra is performed.
For the decays  $D^{*+}\rightarrow D^{0}(K^-\pi^+)\pi^+$ and $D^{*+}\rightarrow D^{0}(K^-K^+)\pi^+$,
we perform maximum likelihood fits to the discriminating variable $\delta m = M_{D^*} - M_{D^0}$, where $M_{D^*}$ and
$M_{D^0}$ are the reconstructed $D^*$ and $D^0$ invariant masses respectively.
Approximately 54 million $D^0 \rightarrow K^-\pi^+$ decays, 7.5 million $D^{*+}\rightarrow D^{0}(K^-\pi^+)\pi^+$ and 1.1 million $D^{*+}\rightarrow D^{0}(K^-K^+)\pi^+$ decays are used.
The mass distributions are shown in Fig.~\ref{fig:controlsamples} (a), (b) and (c). The $D^0 \rightarrow K^-\pi^+$ signal component is modeled as the sum of two Gaussian functions with common mean convolved with a function accounting for final state radiation~\cite{Baracchini:2005wp}, on top of an exponential combinatorial background. The $D^{*+}\rightarrow D^{0}(K^-\pi^+)\pi^+$ and $D^{*+}\rightarrow D^{0}(K^-K^+)\pi^+$ signal components are modeled as the sum of two Gaussian functions convolved with a function taking account of the asymmetric shape of the measured distribution~\cite{Aaij:2011in}. The background is described by an empirical function of the form $1 - e^{-(\delta m - \delta m_0)/\xi}$, where $\delta m_0$ and $\xi$ are free parameters. 
Using the current world average of the integrated $C\!P$ asymmetry for the $D^{0} \rightarrow K^-K^+$ decay~\cite{bib:hfagbase} and neglecting $C\!P$ violation in the Cabibbo-favored $D^0 \rightarrow K^-\pi^+$ decay~\cite{Bianco:2003vb}, from the raw yield asymmetries returned by the mass fits we determine $A_\mathrm{I}(K\pi) = (-1.0 \pm 0.2) \times 10^{-2}$ and $A_\mathrm{R}(K\pi) = (-1.8 \pm 0.2) \times 10^{-3}$, where the uncertainties are statistical only.

The possible existence of a $B^{0}-\overline{B}^0$ production asymmetry is studied by reconstructing a sample of $B^{0}\rightarrow J/\psi K^{*0}$ decays.
$C\!P$ violation in $b\rightarrow c\bar{c}s$ transitions, which is predicted in the SM to be at the $10^{-3}$ level~\cite{Hou:2006du}, is neglected.
The raw asymmetry $A_\mathrm{raw}(B^{0}\rightarrow J/\psi K^{*0})$ is determined from an unbinned maximum likelihood fit to the
$J/\psi(\mu^+\mu^-) K^{*0}(K^+\pi^-)$ and $J/\psi(\mu^+\mu^-) \overline{K}^{*0}(K^-\pi^+)$ mass spectra.
The signal mass peak is modeled as the sum of two Gaussian functions with common mean, whereas the combinatorial background is modeled by an exponential.
The data sample contains approximately 25\hspace{0.5mm}400 $B^{0}\rightarrow J/\psi K^{*0}$ decays. The mass distribution is shown in Fig.~\ref{fig:controlsamples} (d).
To determine the production asymmetry we need to correct for the presence of instrumental asymmetries. Once the necessary corrections are applied, we obtain a value for the $B^0$ production asymmetry $A_\mathrm{P}(B^0)=0.010 \pm 0.013$, where the uncertainty is statistical only.

By using the instrumental and production asymmetries, the correction factor to the raw asymmetry $A_\Delta (B^{0}\rightarrow K\pi) = -0.007 \pm 0.006$ is obtained.
Since the $B^0_s$ meson has no valence quarks in common with those of the incident protons, its production asymmetry is expected to be smaller than for the $B^0$, an expectation that is supported by hadronization models as discussed in Ref. \cite{Lambert:2009zz}. Even conservatively assuming a value of the production asymmetry equal to that for the $B^0$, owing to the small value of $\kappa_s$ the effect of $A_\mathrm{P}(B^0_s)$ is negligible, and we find $A_\Delta (B^{0}_s\rightarrow K \pi) = 0.010 \pm 0.002$.

The systematic uncertainties on the asymmetries fall into the following main categories, related to: (a) PID calibration; (b) modeling of the signal and background components in the maximum likelihood fits; and (c) instrumental and production asymmetries.
Knowledge of PID efficiencies is necessary in this analysis to compute the number of cross-feed background events affecting the mass fit of the $B^0 \rightarrow K\pi$ and $B^0_s \rightarrow K\pi$ decay channels. In order to estimate the impact of imperfect PID calibration, we perform unbinned maximum likelihood fits after having altered the number of cross-feed background events present in the relevant mass spectra according to the systematic uncertainties affecting the PID efficiencies.
An estimate of the uncertainty due to possible imperfections in the description of the final state radiation is determined by varying, over a wide range, the amount of emitted
radiation~\cite{Baracchini:2005wp} in the signal lineshape parameterization.  The possibility of an incorrect description of the core distribution in the signal mass model is investigated by replacing the single Gaussian with the sum of two Gaussian functions with a common mean.  The impact of additional three-body $B$ decays in the $K\pi$ spectrum, not accounted for in the baseline fit --- namely $B\rightarrow \pi\pi\pi$ where one pion is missed in the reconstruction and another is misidentified as a kaon --- is investigated. The mass lineshape of this background component is determined from Monte Carlo simulations, and then the fit is repeated after having modified the baseline parameterization accordingly.
For the modeling of the combinatorial background component, the fit is repeated using a first-order polynomial. Finally, for the case of the cross-feed backgrounds, two distinct systematic uncertainties are estimated: one due to a relative bias in the mass scale of the simulated distributions with respect to the signal distributions in data, and another accounting for the difference in mass resolution between simulation and data. All the shifts from the relevant baseline values are accounted for as systematic uncertainties.
Differences in the kinematic properties of $B$ decays with respect to the charm control samples, as well as different triggers and offline selections, are taken into account by introducing a systematic uncertainty on the values of the $A_\Delta$ corrections. This uncertainty dominates the total systematic uncertainty related to the instrumental and production asymmetries, and can be reduced in future measurements with a better understanding of the dependence of such asymmetries on the kinematics of selected signal and control samples.
The systematic uncertainties for $A_{C\!P}(B^0 \rightarrow K \pi)$ and $A_{C\!P}(B^0_s \rightarrow K\pi)$ are summarized in Table~\ref{tab:SystsumB}.

  \begin{table}
    \caption{Summary of systematic uncertainties on $A_{C\!P}(B^0 \rightarrow K \pi)$ and $A_{C\!P}(B^0_s \rightarrow K\pi)$. The categories (a), (b) and (c) defined in the text are also indicated. The total systematic uncertainties given in the last row are obtained by summing the individual contributions in quadrature.}\label{tab:SystsumB}
    \begin{center}
\resizebox{0.48\textwidth}{!}{
      \begin{tabular}{lcc}
        \hline
\hline
Systematic uncertainty  & $A_{C\!P}(B^0 \rightarrow K \pi)$ \bigstrut & $A_{C\!P}(B^0_s \rightarrow K\pi)$ \\
\hline
$^\mathrm{ (a) }$PID calibration                                           & 0.0012 & 0.001 \\
$^\mathrm{ (b) }$Final state radiation                                   & 0.0026 & 0.010 \\
$^\mathrm{ (b) }$Signal model                                              & 0.0004 & 0.005 \\
$^\mathrm{ (b) }$Combinatorial background             & 0.0001 & 0.009 \\
$^\mathrm{ (b) }$3-body background                       & 0.0009 & 0.007 \\
$^\mathrm{ (b) }$Cross-feed background        & 0.0011 & 0.008 \\
$^\mathrm{ (c) }$Instr. and prod. asym. ($A_\Delta$) & 0.0078   & 0.005  \\
\hline 
Total                                                           & 0.0084  & 0.019 \\
\hline\hline
      \end{tabular}
}
    \end{center}
\vspace{-0.3cm}
  \end{table}

In conclusion we obtain the following measurements of the $C\!P$ asymmetries:
\begin{equation}
A_{C\!P}(B^0 \rightarrow K\pi)=-0.088 \pm 0.011\,\mathrm{(stat)} \pm 0.008\,\mathrm{(syst)}\nonumber
\end{equation}
and
\begin{equation}
A_{C\!P}(B^0_s \rightarrow K\pi)=0.27 \pm 0.08\, \mathrm{(stat)}\pm 0.02\,\mathrm{(syst)}.\nonumber
\end{equation}
The result for $A_{C\!P}(B^0 \rightarrow K\pi)$ constitutes the most precise measurement available to date. It is in good agreement with the current world average provided by the Heavy Flavor Averaging Group $A_{C\!P}(B^0 \rightarrow K\pi)=-0.098^{+0.012}_{-0.011}$~\cite{bib:hfagbase}.  Dividing the central value of $A_{C\!P}(B^0 \rightarrow K\pi)$ by the sum in quadrature of the statistical and systematic uncertainties,
the significance of the measured deviation from zero exceeds $6\sigma$, making this the first observation (greater than 5$\sigma$) of $C\!P$ violation in the $B$ meson sector at a hadron collider.  The same significance computed for $A_{C\!P}(B^0_s \rightarrow K\pi)$ is 3.3$\sigma$, therefore this is the first evidence for $C\!P$ violation in the decays of $B^0_s$ mesons. The result for $A_{C\!P}(B^0_s \rightarrow K\pi)$ is in agreement with the only measurement previously available~\cite{Aaltonen:2011qt}.

\input{acknowledgments}

%% file: acknowledgments.tex
\section*{Acknowledgements}

\noindent We express our gratitude to our colleagues in the CERN accelerator
departments for the excellent performance of the LHC. We thank the
technical and administrative staff at CERN and at the LHCb institutes,
and acknowledge support from the National Agencies: CAPES, CNPq,
FAPERJ and FINEP (Brazil); CERN; NSFC (China); CNRS/IN2P3 (France);
BMBF, DFG, HGF and MPG (Germany); SFI (Ireland); INFN (Italy); FOM and
NWO (The Netherlands); SCSR (Poland); ANCS (Romania); MinES of Russia and
Rosatom (Russia); MICINN, XuntaGal and GENCAT (Spain); SNSF and SER
(Switzerland); NAS Ukraine (Ukraine); STFC (United Kingdom); NSF
(USA). We also acknowledge the support received from the ERC under FP7
and the Region Auvergne.